\documentclass[a4paper]{article}

\usepackage{itgspeech2021}    
\usepackage{times}            
\usepackage[english]{babel}   
\usepackage[ansinew]{inputenc}
\usepackage[T1]{fontenc}      
\usepackage[sort&compress,numbers]{natbib}	
\usepackage{amsmath,amssymb,mathtools,amsfonts}
\usepackage{siunitx}
\usepackage{graphicx}

\usepackage{units}

\usepackage{tikz,pgfplots}
\usetikzlibrary{matrix}
\pgfplotsset{compat=newest, width=3.5cm, height=3.5cm, scale only axis}
\pgfplotsset{every mark/.append style={solid}}
\usepackage{pgfkeys}
    \newenvironment{customlegend}[1][]{%
        \begingroup
        \csname pgfplots@init@cleared@structures\endcsname
        \pgfplotsset{#1}%
    }{%
        \csname pgfplots@createlegend\endcsname
        \endgroup
    }%
    \def\addlegendimage{\csname pgfplots@addlegendimage\endcsname}

\DeclareMathOperator*{\argmin}{argmin}
\DeclarePairedDelimiter\abs{\lvert}{\rvert}%
\DeclarePairedDelimiter\norm{\lVert}{\rVert}%
\makeatletter
\let\oldabs\abs
\def\abs{\@ifstar{\oldabs}{\oldabs*}}
\let\oldnorm\norm
\def\norm{\@ifstar{\oldnorm}{\oldnorm*}}
\makeatother

\usepackage{booktabs}

\usepackage{flushend}

\newcommand\blfootnote[1]{%
  \begingroup
  \renewcommand\thefootnote{}\footnote{#1}%
  \addtocounter{footnote}{-1}%
  \endgroup
}

\usepackage{glossaries}

\newacronym{STFT}{STFT}{short-time Fourier transform}
\newacronym{CTF}{CTF}{convolutive transfer function}
\newacronym{MTF}{MTF}{multiplicative transfer function}
\newacronym{RTF}{RTF}{relative transfer function}
\newacronym{MCLP}{MCLP}{multi channel linear prediction}
\newacronym{MPDR}{MPDR}{minimum power distortionless response}
\newacronym{MVDR}{MVDR}{minimum variance distortionless response}
\newacronym{LCMV}{LCMV}{linear constrained minimum variance}
\newacronym{WPD}{WPD}{weighted power minimization distortionless response}
\newacronym{WPE}{WPE}{weighted prediction error}
\newacronym{TVG}{TVG}{time-varying complex circular Gaussian}
\newacronym{MISO}{MISO}{multiple-input single-output}
\newacronym{MIMO}{MIMO}{multiple-input multiple-output}
\newacronym{SPP}{SPP}{speech presence probability}
\newacronym{PESQ}{PESQ}{perceptual evaluation of speech quality}
\newacronym{FWSSNR}{FWSSNR}{frequency-weighted segmental signal-to-noise ratio}
\newacronym{CW}{CW}{covariance whitening}
\newacronym{VAD}{VAD}{voice activity detection}
\newacronym{PSD}{PSD}{power spectral density}
\newacronym{UCB}{UCB}{unified convolutional beamformer}
\newacronym{IRLS}{IRLS}{iteratively reweighted least squares}

\usepackage{hyperref}
\hypersetup{colorlinks=false,pdfborder={0 0 0}}

\title{Joint Multi-Channel Dereverberation and Noise Reduction \newline Using a Unified Convolutional Beamformer With Sparse Priors}

\author{Henri Gode, Marvin Tammen, Simon Doclo}

\address{Department of Medical Physics and Acoustics and Cluster of Excellence Hearing4all, University of Oldenburg, Germany\\
  Email: \texttt{\{henri.gode,marvin.tammen,simon.doclo\}@uni-oldenburg.de}}

\begin{document}
\sloppy
\maketitle

\begin{abstract}
Recently, the convolutional \gls{WPD} beamformer was proposed, which unifies multi-channel \glsentrylong{WPE} dereverberation and \glsentrylong{MPDR} beamforming. To optimize the convolutional filter, the desired speech component is modeled with a time-varying Gaussian model, which promotes the sparsity of the desired speech component in the \glsentrylong{STFT} domain compared to the noisy microphone signals. In this paper we generalize the convolutional \gls{WPD} beamformer by using an $\ell_p$-norm cost function, introducing an adjustable shape parameter which enables to control the sparsity of the desired speech component. Experiments based on the \textsc{Reverb} challenge dataset show that the proposed method outperforms the conventional convolutional \gls{WPD} beamformer in terms of objective speech quality metrics.
\end{abstract}

\blfootnote{This work was funded by the Deutsche Forschungsgemeinschaft (DFG, German Research Foundation) -- Project ID 390895286 -- EXC 2177/1.}

\glsresetall

\section{Introduction}
In many hands-free speech communication systems such as hearing aids, mobile phones and smart speakers, reverberation and ambient noise may degrade the speech quality and intelligibility of the recorded microphone signals. Reverberation is caused by reflections of a speech source arriving delayed and attenuated at the microphones \cite{kuttruff_room_2016}. Note that early reflections, which arrive roughly in the first $\SI{50}{\milli\s}$ after the direct component, are usually beneficial for human and automatic speech recognition, whereas late reverberation can be detrimental \cite{kuttruff_room_2016, bradley_importance_2003, yoshioka_making_2012, warzybok_effects_2013}. In many scenarios the microphones also capture undesired noise, e.g., originating from traffic, house appliances or industrial machinery.

First, to achieve noise reduction, a commonly used multi-microphone noise reduction technique is the \gls{MPDR} beamformer \cite{ cox_resolving_1973, veen_beamforming_1988, trees_optimum_2004, doclo_multichannel_2015}, which aims at minimizing the output power while leaving the desired speech component undistorted. To implement the \gls{MPDR} beamformer, the \gls{RTF} vector of the desired speech source is required, which can be estimated, e.g., using the covariance whitening method, assuming that an estimate of the noise covariance matrix is available \cite{markovich_multichannel_2009, serizel_low-rank_2014, markovich-golan_performance_2018}.

Second, to achieve dereverberation, the so-called \gls{WPE} technique is commonly applied in the \gls{STFT} domain \cite{nakatani_blind_2008, nakatani_speech_2010, yoshioka_generalization_2012}. It uses a convolutional filter, to estimate the late reverberation component by modeling the desired speech component with a \gls{TVG} model. The convolutional filter is applied to a number of past \gls{STFT} frames excluding a few most recent frames, with the aim of preserving the early reflections. Since anechoic speech is sparser than reverberant speech in the \gls{STFT} domain, a variant of \gls{WPE} with sparse priors has been proposed in \cite{jukic_multi-channel_2015, jukic_group_2015, jukic_adaptive_2017}, which uses an $\ell_p$-norm cost function to optimize the convolutional filter. Since both cost functions do not have analytic solutions, it has been proposed to use iterative alternating optimization schemes, such as the \gls{IRLS} method \cite{chartrand_iteratively_2008, rao_affine_1999, jukic_multi-channel_2015}.

Aiming at joint dereverberation and noise reduction, it was proposed to perform \gls{WPE} as a preprocessing stage before \gls{MPDR} beamforming in a combined cascade system \cite{delcroix_strategies_2015, yang_dereverberation_2018}. The so-called \gls{WPD} convolutional beamformer proposed in \cite{nakatani_unified_2019, nakatani_simultaneous_2019, nakatani_maximum_2019, boeddeker_jointly_2020} was shown to outperform those cascade systems by unifying the optimization of the convolutional \gls{WPE} filter and the \gls{MPDR} beamformer. The unified convolutional \gls{WPD} beamformer 
is optimized similarly to the convolutional \gls{WPE} filter by modeling the desired speech component with a \gls{TVG} model and additionally introducing a distortionless constraint using the \glspl{RTF} of the desired speech source. 

In this paper we propose to optimize the convolutional beamformer coefficients by explicitly taking into account that the desired speech component is sparser than the noisy reverberant speech in the \gls{STFT} domain. 
Hence, similar to the \gls{WPE} variant in \cite{jukic_multi-channel_2015, jukic_group_2015}, we propose to optimize the convolutional beamformer coefficients using an $\ell_p$-norm cost function with an additional distortionless constraint. The optimization is performed using the \gls{IRLS} method. 
We evaluate the influence of the shape parameter $p$ of the $\ell_p$-norm cost function and the influence of initialization in terms of \gls{PESQ} and \gls{FWSSNR} \cite{rix_perceptual_2001, hu_evaluation_2008}. 
The simulation results show that the speech enhancement performance 
can be improved by setting the shape parameter $p$ to an appropriate value. In addition the results show that the multi-channel initialization approach results in a faster convergence of the iterative optimization scheme than single-channel initialization.

\section{Signal Model}
\label{sec:SignalModel}
We consider a single speech source captured by $M$ microphones in a noisy and reverberant acoustic environment. The \gls{STFT} coefficients of the microphone signals at time frame $t$ and any frequency bin are denoted as
\begin{align}
    \mathbf{y}_t = \left[ y_{1,t} \enspace\ldots\enspace y_{M,t} \right]^\mathrm{T} \in \mathbb{C}^{M\times 1},
\end{align}
with $\left(\cdot\right)^{\mathrm{T}}$ denoting the transpose operator. The frequency index is omitted for brevity since it is assumed that each frequency subband is independent and can hence be processed individually. Assuming that $T$ time frames are available, the batch matrix of the microphone signals is defined as
\begin{align}
    \mathbf{Y} = \left[ \mathbf{y}_{1} \enspace\ldots\enspace \mathbf{y}_{T} \right] \in \mathbb{C}^{M\times T}.
\end{align}
As in \cite{nakatani_blind_2008, nakatani_speech_2010, yoshioka_generalization_2012, jukic_multi-channel_2015, jukic_group_2015} the multi-channel microphone signal $\mathbf{y}_t$ is modeled as the convolution of the clean speech signal $s_t$ with the stationary multi-channel \gls{CTF} matrix $\mathbf{A} = \left[ \mathbf{a}_{0} \enspace\ldots\enspace \mathbf{a}_{L_a-1} \right] \in \mathbb{C}^{M\times L_a}$ plus additive noise $\mathbf{n}_t \in \mathbb{C}^{M\times 1}$, i.e.
\begin{align}
    \mathbf{y}_t = \sum_{l=0}^{L_a-1} \mathbf{a}_l s_{t-l} + \mathbf{n}_t = \underbrace{\sum_{l=0}^{\tau-1} \mathbf{a}_l s_{t-l}}_{\coloneqq\mathbf{d}_t} + \underbrace{\sum_{l=\tau}^{L_a-1} \mathbf{a}_l s_{t-l}}_{\coloneqq\mathbf{r}_t} + \mathbf{n}_t,
\end{align}
where $L_a$ denotes the number of taps of the \glspl{CTF} and $\tau$ denotes the so-called prediction delay. This delay separates the early reflections from the late reverberation, i.e. the reverberant speech is decomposed into the desired speech component $\mathbf{d}_t \in \mathbb{C}^{M\times 1}$
and the late reverberation component $\mathbf{r}_t \in \mathbb{C}^{M\times 1}$. 
The desired speech component can be approximated using the stationary \gls{MTF} vector $\mathbf{v} \in \mathbb{C}^{M\times 1}$ as \cite{avargel_multiplicative_2007}
\begin{align}
    \mathbf{d}_t \approx \mathbf{v} s_t = \mathbf{\Tilde{v}}_m d_{m,t}\quad \mathrm{with}\quad m \in \{1,...,M\}, 
    \label{eq:dsigmodel}
\end{align}
where $d_{m,t}$ and $\mathbf{d}_{m} \in \mathbb{C}^{1\times T}$ denote the desired speech component in the reference microphone $m$ at time frame $t$ and the full batch vector, respectively. The vector $\mathbf{\Tilde{v}}_m = \mathbf{v} / v_m \in \mathbb{C}^{M\times 1}$ denotes the \gls{RTF} vector, where $v_m$ is the $m$-th entry of $\mathbf{v}$.

\subsection{Estimating RTF vector by Covariance Whitening}
As proposed in \cite{markovich_multichannel_2009, serizel_low-rank_2014, markovich-golan_performance_2018}, the \gls{RTF} vector $\mathbf{\Tilde{v}}_m$ can be estimated with the covariance whitening method, assuming that $\mathbf{d}_t$ and $\mathbf{n}_t$ are uncorrelated and that $\mathbf{r}_t \approx \mathbf{0}$. The noisy covariance matrix $\mathbf{R}_y= \nicefrac{1}{T}\sum_{t=1}^{T}\mathbf{y}_t\mathbf{y}_t^{\mathrm{H}}$ can be decomposed into the speech covariance matrix $\mathbf{R}_d= \nicefrac{1}{T}\sum_{t=1}^{T}\mathbf{d}_t\mathbf{d}_t^{\mathrm{H}}$ and the noise covariance matrix $\mathbf{R}_n= \nicefrac{1}{T}\sum_{t=1}^{T}\mathbf{n}_t\mathbf{n}_t^{\mathrm{H}}$ with $\left(\cdot\right)^{\mathrm{H}}$ denoting the Hermitian operator, i.e.
\begin{align}
    \mathbf{R}_y = \mathbf{R}_d + \mathbf{R}_n \approx \phi_s \mathbf{v}\mathbf{v}^{\mathrm{H}}+\mathbf{R}_n,
\end{align}
where $\phi_s$ denotes the \gls{PSD} of the speech component, and the \gls{MTF} approximation in \eqref{eq:dsigmodel} has been used for the speech covariance matrix $\mathbf{R}_d$. Assuming that the (positive definite) noise covariance matrix is available, the noisy covariance matrix can be whitened as
\begin{align}
    \mathbf{R}_n^{\nicefrac{-\mathrm{H}}{2}}\mathbf{R}_y\mathbf{R}_n^{\nicefrac{-1}{2}} &= \mathbf{R}_n^{\nicefrac{-\mathrm{H}}{2}}\mathbf{R}_d\mathbf{R}_n^{\nicefrac{-1}{2}} + \mathbf{I} \\ &\approx \phi_s \mathbf{R}_n^{\nicefrac{-\mathrm{H}}{2}}\mathbf{v}\mathbf{v}^{\mathrm{H}}\mathbf{R}_n^{\nicefrac{-1}{2}}+\mathbf{I}
\end{align}
where $\mathbf{I}$ denotes the identity matrix and $\mathbf{R}_n^{\nicefrac{1}{2}}$ is any matrix square root of $\mathbf{R}_n$ so that $\mathbf{R}_n^{\nicefrac{\mathrm{H}}{2}}\mathbf{R}_n^{\nicefrac{1}{2}} = \mathbf{R}_n$. The principal eigenvector $\mathbf{\dot{v}}$ of $\mathbf{R}_n^{\nicefrac{-\mathrm{H}}{2}}\mathbf{R}_y\mathbf{R}_n^{\nicefrac{-1}{2}}$ is equal to $\alpha \mathbf{R}_n^{\nicefrac{-\mathrm{H}}{2}}\mathbf{v}$, where $\alpha \neq 0$ denotes an arbitrary scaling factor. The \gls{RTF} vector $\mathbf{\Tilde{v}}_m$ can be obtained by de-whitening $\mathbf{\dot{v}}$ and normalizing w.r.t its $m$-th entry, i.e.
\begin{align}
    \mathbf{\Tilde{v}}_m = \frac{\mathbf{v}}{v_m} = \frac{\mathbf{R}_n^{\nicefrac{H}{2}}\mathbf{\dot{v}}}{\mathbf{e}_m^{\mathrm{T}}\mathbf{R}_n^{\nicefrac{H}{2}}\mathbf{\dot{v}}} = \frac{\mathbf{R}_n^{\nicefrac{H}{2}}\mathbf{R}_n^{\nicefrac{-\mathrm{H}}{2}}\mathbf{v}}{\mathbf{e}_m^{\mathrm{T}}\mathbf{R}_n^{\nicefrac{H}{2}}\mathbf{R}_n^{\nicefrac{-\mathrm{H}}{2}}\mathbf{v}}
\end{align}
where $\mathbf{e}_m$ denotes a selection vector with the $m$-th entry equal to one and all other entries equal to zero.

\subsection{Convolutional Filter}
To obtain an estimate $z_{m,t}$ of the desired speech component $d_{m,t}$ in the reference microphone $m$ at time frame $t$ a convolutional filter $\mathbf{\Bar{h}}_{m}\in\mathbb{C}^{M\left(L_{h}-\tau+1\right)\times 1}$, can be applied to the noisy \gls{STFT} vector, i.e. \cite{nakatani_blind_2008, nakatani_speech_2010, yoshioka_generalization_2012, jukic_multi-channel_2015, jukic_group_2015, nakatani_unified_2019, nakatani_simultaneous_2019, nakatani_maximum_2019, boeddeker_jointly_2020}
\begin{align}
    z_{m,t} = {\mathbf{\Bar{h}}_{m}}^{\mathrm{H}}\mathbf{\Bar{y}}_{t},
    \label{eq:WPD_Filter1}
\end{align}
where the stacked microphone signal vector $\mathbf{\Bar{y}}_{t}$ is defined as
\begin{align}
    \mathbf{\Bar{y}}_{t} &= \left[ \mathbf{y}^{\mathrm{T}}_{t} \enspace\mathbf{y}^{\mathrm{T}}_{t-\tau} \enspace\ldots\enspace \mathbf{y}^{\mathrm{T}}_{t-L_{h}+1} \right]^{\mathrm{T}} \in\mathbb{C}^{M\left(L_{h}-\tau+1\right)\times 1}.
\end{align}
Note that the vector $\mathbf{\Bar{y}}_{t}$ only includes a subset of the $L_h$ most recent frames, i.e. it includes the current frame but excludes $\tau-1$ frames, aiming at preserving the early reflections. The batch vector $\mathbf{z}_{m} \in \mathbb{C}^{1 \times T}$ containing estimates of the desired speech component for all time frames can be obtained as 
\begin{align}
    \mathbf{z}_{m} = {\mathbf{\Bar{h}}_{m}}^{\mathrm{H}}\mathbf{\Bar{Y}},
    \label{eq:WPD_Filter}
\end{align}
with
\begin{align}
    \mathbf{\Bar{Y}} &= \left[ \mathbf{\Bar{y}}_{1} \enspace\ldots\enspace \mathbf{\Bar{y}}_{T} \right] \in \mathbb{C}^{M\left(L_{h}-\tau+1\right)\times T}.
\end{align}
\section{Conventional WPD \newline using TVG model}

\label{sec:conventional}
In \cite{nakatani_unified_2019, nakatani_maximum_2019, boeddeker_jointly_2020}, the \gls{WPD} convolutional beamformer has been proposed to achieve joint dereverberation and noise reduction. The \gls{WPD} convolutional beamformer $\mathbf{\Bar{h}}_{m}$ is optimized by modeling the desired speech component $d_{m,t}$ in the reference microphone $m$ with a \gls{TVG} model similarly to \gls{WPE} dereverberation \cite{nakatani_blind_2008, nakatani_speech_2010, jukic_multi-channel_2015} and additionally introducing a distortionless constraint similarly to the \gls{MPDR} beamformer \cite{veen_beamforming_1988}. The corresponding negative log-likelihood $\mathcal{L}$ to be minimized is given by \cite{nakatani_maximum_2019} 
\begin{align}
    \label{eq:WPD_CostFun1}
    \mathcal{L}\left(\mathbf{\Bar{h}}_{m}, \mathbf{\Lambda}\right) &= \frac{1}{T}\sum_{t=1}^{T} \left(  \ln{\lambda_t}+\frac{\abs{z_{m,t}}^2}{\lambda_t}\right)\\
    &= \frac{1}{T} \left( \mathrm{tr}\left(\ln{\mathbf{\Lambda}}\right)+\mathbf{z}_{m}\mathbf{\Lambda}^{-1}\mathbf{z}_{m}^{\mathrm{H}} \right),
    \label{eq:WPD_CostFun2}
\end{align}
where $\mathrm{tr}\left(\cdot\right)$ denotes the trace operator, $\lambda_t = \mathbb{E}\left[\abs{d_{m,t}}^2\right]$ denotes the \gls{PSD} of the desired speech component at frame $t$, corresponding to the time-varying variance of the \gls{TVG} model, and $\mathbf{\Lambda} \in \mathbb{R}_{+}^{T \times T}$ denotes a diagonal matrix containing these variances for all $T$ time frames. The distortionless constraint is given by \cite{veen_beamforming_1988}
\begin{align}
    \mathbf{\Bar{h}}_{m}^{\mathrm{H}}\mathbf{\Bar{v}}_{m} = 1,
    \label{eq:dist_constraint}
\end{align}
where $\mathbf{\Bar{v}}_{m} = \begin{bmatrix} \mathbf{\Tilde{v}}_{m}^{\mathrm{T}} & \mathbf{0}^{\mathrm{T}} \end{bmatrix}^{\mathrm{T}}$ and $\mathbf{0}$ is a vector containing $M\left(L_h-\tau\right)$ zeros. Note that the cost function in \eqref{eq:WPD_CostFun2} depends on the \glspl{PSD} of the desired speech component, which are obviously not available in practice. Since the cost function is non-convex and does not have an analytic solution it has been proposed in \cite{nakatani_unified_2019, nakatani_maximum_2019} to use an iterative alternating optimization scheme to approximate the optimal filter. In the first of the two alternating optimization steps, the variances $\mathbf{\Lambda}$ are fixed to optimize the convolutional filter, and in the second step the convolutional filter is fixed to update the variances using the estimate of the desired speech component.

\paragraph{(1) Estimating the filter by fixing the variances}
\label{sec:_Est_Filter}
By fixing the variances $\mathbf{\Lambda}_i$ in the $i$-th iteration of the alternating optimization and using \eqref{eq:WPD_Filter}, the cost function in \eqref{eq:WPD_CostFun2} to be minimized reduces to 
\begin{align}
    \label{eq:WPD_likelihood}
    \mathcal{L}\left(\mathbf{\Bar{h}}_{m,i}\right) &\propto \frac{1}{T}\mathbf{z}_{m,i}\mathbf{\Lambda}_{i}^{-1}\mathbf{z}_{m,i}^{\mathrm{H}} \\ &= \frac{1}{T}\mathbf{\Bar{h}}_{m,i}^{\mathrm{H}}\mathbf{\Bar{Y}}\mathbf{\Lambda}_{i}^{-1}\mathbf{\Bar{Y}}^{\mathrm{H}}\mathbf{\Bar{h}}_{m,i} \\ &= \mathbf{\Bar{h}}_{m,i}^{\mathrm{H}}\mathbf{\Bar{R}}_{y,i}\mathbf{\Bar{h}}_{m,i},
    \label{eq:likelihood_Ry}
\end{align}
where $\mathbf{\Bar{R}}_{y,i} = \nicefrac{1}{T}\mathbf{\Bar{Y}}\mathbf{\Lambda}_{i}^{-1}\mathbf{\Bar{Y}}^{\mathrm{H}}$ denotes the power-weighted noisy sample covariance matrix of the stacked microphone signals. The solution of the resulting constrained optimization problem
\begin{align}
    \label{eq:constraint_opt}
    \mathbf{\Bar{h}}_{m,i}^{\mathrm{opt}} = \argmin_{\mathbf{\Bar{h}}_{m,i}} \left( \mathbf{\Bar{h}}_{m,i}^{\mathrm{H}} \mathbf{\Bar{R}}_{y,i} \mathbf{\Bar{h}}_{m,i}\right) \ \ \mathrm{s.t.}\ \ \mathbf{\Bar{h}}_{m,i}^{\mathrm{H}}\mathbf{\Bar{v}}_{m} = 1
\end{align}
is given by the \gls{MPDR} beamformer \cite{cox_resolving_1973}:
\begin{align}
    \mathbf{\Bar{h}}_{m,i}^{\mathrm{opt}} = \frac{\mathbf{\Bar{R}}_{y,i}^{-1} \mathbf{\Bar{v}}_{m}}{\mathbf{\Bar{v}}_{m}^{\mathrm{H}} \mathbf{\Bar{R}}_{y,i}^{-1} \mathbf{\Bar{v}}_{m}}.
    \label{eq:MISOWPD_Beamformer}
\end{align}

\paragraph{(2) Estimating the variances by fixing the filter}
\label{sec:_Est_Variances}
By now fixing the convolutional filter $\mathbf{\Bar{h}}_{m,i}$, the variances in the $i$-th iteration can be updated by minimizing \eqref{eq:WPD_CostFun1} \cite{nakatani_speech_2010, jukic_multi-channel_2015}, i.e.
\begin{align}
    \lambda_{t,i+1} = \abs{z_{m,t,i}}^2 = \abs{{\mathbf{\Bar{h}}_{m,i}}^{\mathrm{opt,H}}\mathbf{\Bar{y}}_{t}}^2.
    \label{eq:update_variances}
\end{align}

\section{Proposed Method \newline using Sparse Priors}
We propose to optimize the convolutional beamformer coefficients by explicitly taking into account that the desired speech component is sparser than the noisy reverberant speech in the \gls{STFT} domain. 
Hence, instead of the \gls{TVG} model in \eqref{eq:WPD_CostFun1}, we propose to optimize the convolutional filter in \eqref{eq:WPD_Filter} using an $\ell_p$-norm cost function similarly to the \gls{WPE} variant in \cite{jukic_multi-channel_2015, jukic_group_2015}, i.e.
\begin{align}
    \mathcal{L}\left(\mathbf{\Bar{h}}_{m}\right) &\propto \norm{\mathbf{z}_{m}}_{p}^{p} \propto \frac{1}{T}\sum_{t=1}^{T} \abs{z_{m,t}}^p,
    \label{eq:WPD_lp_norm_CostFun}
\end{align}
where $p \in (0,2] $ denotes the so-called shape parameter. The shape parameter determines the sparsity of the cost function, where small values of $p$ promote sparsity. It should be noted that for $0 < p < 1$ this cost function is non-convex. In addition, we use the same distortionless constraint $\mathbf{\Bar{h}}_{m}^{\mathrm{H}}\mathbf{\Bar{v}}_{m} = 1$ as for the conventional \gls{WPD} beamformer in \eqref{eq:dist_constraint}. Similarly as in \cite{rao_affine_1999, jukic_multi-channel_2015}, we propose to use an \gls{IRLS} method with the basic idea to replace the non-convex $\ell_p$-norm minimization problem with a series of convex $\ell_2$-norm minimization subproblems. In each iteration, the $\ell_2$-norm minimization subproblem has an analytic solution, which modifies the optimization problem of the next iteration. This leads to an iterative alternating optimization scheme similar to the optimization scheme for \gls{WPD} in  Section~\ref{sec:conventional}. The two alternating steps are described in the following paragraphs.

\paragraph{(1) Constrained \ensuremath{\ell_2}--Norm Subproblem Minimization}
\label{sec:_l2_norm_subproblem}
In each iteration $i$, the non-convex cost function in \eqref{eq:WPD_lp_norm_CostFun} is replaced with a convex weighted $\ell_2$-norm cost function, i.e.
\begin{align}
    \mathcal{L}\left(\mathbf{\Bar{h}}_{m,i}\right) &\propto \frac{1}{T}\mathbf{z}_{m,i}\mathbf{W}_{i}\mathbf{z}_{m,i}^{\mathrm{H}},
    \label{eq:WPD_l2_norm_CostFun1}
\end{align}
where $\mathbf{W}_i$ denotes the diagonal weighting matrix, i.e.
\begin{align}
    \mathbf{W}_i = \mathrm{diag}\left(\left[w_{1,i} \enspace\ldots\enspace w_{T,i} \right]^\mathrm{T}\right) \in \mathbb{R}_{+}^{T \times T},
    \label{eq:weights}
\end{align}
where the weights $w_{t,i}$ are real-valued and positive. It should be noted that the cost function in \eqref{eq:WPD_l2_norm_CostFun1} is similar to \eqref{eq:WPD_likelihood}, where the weight matrix $\mathbf{W}_i$ takes the role of $\mathbf{\Lambda}_i^{-1}$. Hence, similarly to \eqref{eq:MISOWPD_Beamformer}, the solution minimizing \eqref{eq:WPD_l2_norm_CostFun1} subject to the distortionless constraint in \eqref{eq:dist_constraint} is equal to
\begin{align}
    \mathbf{\Bar{h}}_{m,i}^{\mathrm{opt}} = \frac{\left(\mathbf{\Bar{R}}^{\mathbf{W}}_{y,i}\right)^{-1} \mathbf{\Bar{v}}_{m}}{\mathbf{\Bar{v}}_{m}^{\mathrm{H}} \left(\mathbf{\Bar{R}}^{\mathbf{W}}_{y,i}\right)^{-1} \mathbf{\Bar{v}}_{m}}.
    \label{eq:l2_WPD_Beamformer}
\end{align}
where $\mathbf{\Bar{R}}_{y,i}^{\mathbf{W}} = \nicefrac{1}{T}\mathbf{\Bar{Y}}\mathbf{W}_{i}\mathbf{\Bar{Y}}^{\mathrm{H}}$ denotes the weighted noisy sample covariance matrix of the stacked microphone signals.


\paragraph{(2) Updating the Weights}
\label{sec:_EST_Weights}
Similarly as in \cite{rao_affine_1999, jukic_multi-channel_2015}, in each iteration the weights in \eqref{eq:weights} are updated as
\begin{align}
    w_{t,i+1} = \frac{1}{\abs{z_{m,t,i}}^{2-p}} = \frac{1}{\abs{{\mathbf{\Bar{h}}_{m,i}}^{\mathrm{opt,H}}\mathbf{\Bar{y}}_{t}}^{2-p}},
    \label{eq:update_weights}
\end{align}
so that \eqref{eq:WPD_l2_norm_CostFun1} is a first-order approximation of \eqref{eq:WPD_lp_norm_CostFun}. It should be noted that for $p=0$, the conventional and proposed optimization schemes are equivalent, since $w_{t,i+1} = \lambda_{t,i+1}^{-1}$ yielding $\mathbf{\Bar{R}}_{y,i}^{\mathbf{W}} = \mathbf{\Bar{R}}_{y,i}$. This means that the conventional \gls{WPD} algorithm models the desired speech component as the most sparse, while for larger values of $p$ the desired speech component is modeled less sparse. 

\section{Initialization}
\label{sec:initialization}
Both the conventional \gls{WPD} beamformer and the proposed $\ell_p$-norm \gls{WPD} beamformer are based on an iterative alternating optimization scheme. In each iteration, first the convolutional filter is estimated, based on which the variances or equivalent weights are updated. These updates modify the estimation of the convolutional filter in the next iteration. However, the update equations \eqref{eq:update_variances} and \eqref{eq:update_weights} depend on the estimate of the desired speech component, which is obviously not available in the first iteration. One option to initialize this estimate is to simply use the noisy and reverberant reference microphone signal, i.e.
\begin{align}
    \lambda_{t,1} = \abs{y_{m,t}}^2 \quad\mathrm{and}\quad w_{t,1} = \frac{1}{\abs{y_{m,t}}^{2-p}}.
    \label{eq:MISOinit_weights}
\end{align}
Another option is to use all noisy and reverberant microphone signals, similarly to \cite{jukic_group_2015, drude_integrating_2018}, i.e.
\begin{align}
    \lambda_{t,1} = \frac{\norm{\mathbf{y}_{t}}_2^2}{M}, \quad\quad w_{t,1} = \frac{M}{\norm{\mathbf{y}_{t}}_2^{2-p}}.
    \label{eq:MIMOinit_weights}
\end{align}

\section{Experiments}
\label{sec:Experiments}
In this section, we compare the performance of the conventional \gls{WPD} beamformer with the proposed beamformer. More in particular, we evaluate the influence of the shape parameter $p$ and different initialization approaches.

\subsection{Dataset, Evaluation Metrics and Analysis Conditions}

\begin{table}[t]
\caption{Algorithm Parameters}
\label{table}
\small
\setlength{\tabcolsep}{3pt}
\begin{tabular}{|p{90pt}|p{40pt}|p{80pt}|}
\hline
\textbf{Parameter}& 
\textbf{Symbol}& 
\textbf{Value}
\\ \toprule

frame length        &          & $\SI{512}{taps} \;\widehat{=}\; \SI{32}{\milli\s}$                      \\ \hline
frame shift                                                                            &                        & $\SI{128}{taps} \;\widehat{=}\; \SI{8}{\milli\s}$                       \\ \hline
window                                                                                 &                        & square-root-Hann                     \\ \hline
prediction delay                                                                       & $\tau$            & $\SI{4}{frames} \;\widehat{=}\; \SI{32}{\milli\s}$                         \\ \hline
\begin{tabular}[c]{@{}l@{}}prediction filter length\end{tabular} & $L_{h}$   & $\SI{12}{frames} \;\widehat{=}\; \SI{96}{\milli\s}$                        \\ \hline
reference microphone                                                                      & $m$              & $1$  \\ \hline
\end{tabular}
\end{table}

We used the simulated data of the development set of the \textsc{Reverb} challenge \cite{kinoshita_reverb_2013, kinoshita_summary_2016} with sampling frequency $f_s = \SI{16}{\kilo\Hz}$. The dataset simulates a circular microphone array with 8 channels in six different reverberation conditions resulting from two speaker-to-microphone distances of $\SI{50}{\centi\m}$ and $\SI{200}{\centi\m}$ and three different rooms with reverberation times of $T_{60} \in \{\SI{0.3}{\s}, \SI{0.6}{\s}, \SI{0.7}{\s}\}$. After convolving the clean utterances with one of the six room impulse responses, stationary diffuse background noise was added with a signal-to-noise ratio of $\SI{20}{\dB}$. As objective measures of the speech quality we computed \gls{PESQ} and \gls{FWSSNR} scores \cite{rix_perceptual_2001, hu_evaluation_2008}, where we used the clean speech signal $s_t$ as the reference signal. The parameters of the algorithms are stated in Tab.~\ref{table}. The \gls{RTF} vector $\mathbf{\Tilde{v}}_{m}$ was estimated blindly using the \gls{CW} method \cite{markovich_multichannel_2009, serizel_low-rank_2014, markovich-golan_performance_2018}, assuming that noise-only frames are present in the first $\SI{225}{\milli\s}$ and the last $\SI{75}{\milli\s}$ to estimate the noise covariance matrix $\mathbf{R}_n$.

\subsection{Results}
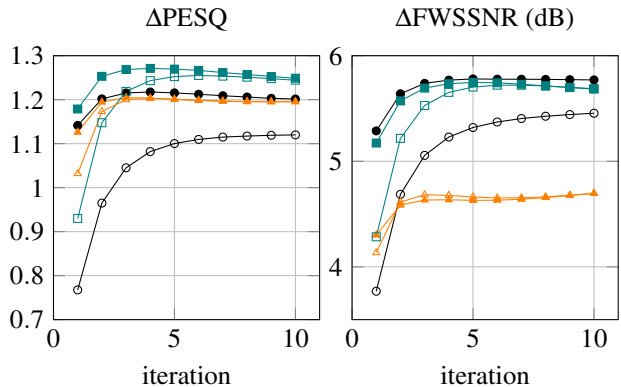
\begin{figure}[t]
    \begin{center}
    \begin{tikzpicture}
\begin{customlegend}[legend columns=4,legend style={align=center,column sep=0.5ex, at={(1,0)}},
        legend entries={MC: ,
                        $p=0$ ,
                        $p=0.5$ ,
                        $p=1$ ,
                        SC: ,
                        $p=0$,
                        $p=0.5$ ,
                        $p=1$ ,
                        }]
        \addlegendimage{empty legend}
        \addlegendimage{mark=*, mark size=1.5pt, black,solid,line legend}
        \addlegendimage{mark=square*, mark size=1.5pt, teal}  
        \addlegendimage{mark=triangle*, mark size=1.5pt, orange}
        \addlegendimage{empty legend}
        \addlegendimage{mark=o, mark size=1.5pt, black,solid}
        \addlegendimage{mark=square, mark size=1.5pt, teal}  
        \addlegendimage{mark=triangle, mark size=1.5pt, orange}
        \end{customlegend}
        \end{tikzpicture}
      \end{center}
    \begin{tikzpicture}
    \matrix{
        \begin{axis}[name=plot7,
                     title={$\Delta$PESQ},
                     xlabel={iteration},
                     xmin = 0,
                     xmax = 11,
                     ymin = 0.7,
                     ymax = 1.3,
                     grid=major,
                     ytick distance=0.1,
                     ]
            \addplot [mark=*, mark size=1.5pt, black]
            coordinates {
     (1,1.1416) (2,1.2018) (3,1.2153) (4,1.2175) (5,1.2155) (6,1.2124) (7,1.209) (8,1.2061) (9,1.2031) (10,1.2015)
     };
            
            \addplot [mark=square*, mark size=1.5pt, teal]
            coordinates {
     (1,1.1788) (2,1.2531) (3,1.2686) (4,1.2711) (5,1.2695) (6,1.2667) (7,1.2617) (8,1.2573) (9,1.2528) (10,1.2488)
     };
     
             \addplot [mark=triangle*, mark size=1.5pt, orange]
             coordinates {
      (1,1.1262) (2,1.1946) (3,1.2049) (4,1.2037) (5,1.2007) (6,1.1982) (7,1.1964) (8,1.1954) (9,1.1953) (10,1.1955)
      };
     
            \addplot [mark=o, mark size=1.5pt, black]
            coordinates {
     (1,0.76764) (2,0.96497) (3,1.0451) (4,1.082) (5,1.1003) (6,1.1097) (7,1.115) (8,1.1173) (9,1.1191) (10,1.12)
     };
     
            \addplot [mark=square, mark size=1.5pt, teal]
            coordinates {
     (1,0.93014) (2,1.1476) (3,1.2189) (4,1.2433) (5,1.2525) (6,1.2551) (7,1.2536) (8,1.2512) (9,1.2478) (10,1.2444)
     };
     
             \addplot [mark=triangle, mark size=1.5pt, orange]
             coordinates {
      (1,1.032) (2,1.173) (3,1.2001) (4,1.203) (5,1.201) (6,1.1988) (7,1.1969) (8,1.1958) (9,1.1953) (10,1.1954)
      };
        \end{axis}
        
        &\begin{axis}[name=plot8,
                     title={$\Delta$FWSSNR (dB)},
                     xlabel={iteration},
                     xmin = 0,
                     xmax = 11,
                     ymin = 3.5,
                     ymax = 6,
                     grid=major,
                     ]
            \addplot [mark=*, mark size=1.5pt, black]
            coordinates {
     (1,5.2879) (2,5.6393) (3,5.7378) (4,5.7689) (5,5.7802) (6,5.7776) (7,5.7771) (8,5.7758) (9,5.7731) (10,5.7707)
     };
            
            \addplot [mark=square*, mark size=1.5pt, teal]
            coordinates {
     (1,5.1728) (2,5.5722) (3,5.6932) (4,5.735) (5,5.7465) (6,5.7434) (7,5.7299) (8,5.7141) (9,5.6976) (10,5.6843)
     };
     
             \addplot [mark=triangle*, mark size=1.5pt, orange]
             coordinates {
      (1,4.3008) (2,4.5848) (3,4.6333) (4,4.6351) (5,4.6298) (6,4.6311) (7,4.6404) (8,4.6562) (9,4.6756) (10,4.6959)
      };
     
            \addplot [mark=o, mark size=1.5pt, black]
            coordinates {
     (1,3.769) (2,4.687) (3,5.0557) (4,5.2296) (5,5.3199) (6,5.3719) (7,5.4049) (8,5.4267) (9,5.4424) (10,5.455)
     };
     
            \addplot [mark=square, mark size=1.5pt, teal]
            coordinates {
     (1,4.2842) (2,5.2167) (3,5.5286) (4,5.6527) (5,5.7028) (6,5.7209) (7,5.7204) (8,5.7131) (9,5.7001) (10,5.6878)
     };
     
             \addplot [mark=triangle, mark size=1.5pt, orange]
             coordinates {
      (1,4.1337) (2,4.6134) (3,4.6832) (4,4.6767) (5,4.6617) (6,4.6527) (7,4.6542) (8,4.6633) (9,4.678) (10,4.6958)
      };
     
        \end{axis}\\
    };
    \end{tikzpicture}

    \caption{Average \gls{PESQ} and \gls{FWSSNR} improvement vs. number of iterations for different shape parameters $p$. Filled markers correspond to multi-channel (MC) initialization of the weights as in \eqref{eq:MIMOinit_weights}, while empty markers correspond to single-channel (SC) initialization of the weights as in \eqref{eq:MISOinit_weights}.}
    \label{fig:Delta_Perf_Lines}
\end{figure}

Fig.~\ref{fig:Delta_Perf_Lines} 
shows the average \gls{PESQ} and \gls{FWSSNR} improvement vs. the number of iterations of the $\ell_p$-norm \gls{WPD} algorithm for different shape parameters $p$ and initializations (see Section~\ref{sec:initialization}). First, the results show that for all considered parameter choices the speech quality is improved in terms of \gls{PESQ} and \gls{FWSSNR} compared to the noisy reference microphone signal. Second, the results after $I=10$ iterations show that for both initializations a shape parameter of $p=0.5$ outperforms the conventional method ($p=0$), which stronger promotes sparsity, and its variant ($p=1$), which promotes sparsity less, in terms of \gls{PESQ} and \gls{FWSSNR} improvement, except for the \gls{FWSSNR} improvement of the conventional method for the multi-channel initialization. Third, it can be observed that the multi-channel initialization consistently outperforms the single-channel initialization in terms of convergence speed and for the conventional method ($p=0$) also in terms of performance after $I=10$ iterations. These results show the influence of the shape parameter $p$ and the initialization on the performance of the proposed \gls{WPD} beamformer with sparse priors.

\section{Conclusion}

In this paper we proposed a novel convolutional beamformer for joint dereverberation and noise reduction, based on a sparse prior for modeling the desired speech component. The proposed $\ell_p$-norm \gls{WPD} beamformer can be interpreted as a generalization of the conventional \gls{WPD} beamformer using the \gls{TVG} model. We propose to compute the convolutional beamformer using an \gls{IRLS} method, where the non-convex constrained $\ell_p$-norm minimization problem is replaced with a series of convex constrained $\ell_2$-norm minimization subproblems. The experimental results show that speech enhancement performance can be consistently improved by setting the shape parameter $p$ to an appropriate value. In addition, the results show that multi-channel initialization improves the performance and the convergence speed.

\small
\bibliographystyle{ieeetr}
\bibliography{example}


\end{document}